\documentclass[%
onecolumn,
amsmath,amssymb,
aps,
]{revtex4-2}
\usepackage[exponent-product = \cdot]{siunitx}
\usepackage[nolist,nohyperlinks]{acronym}
\usepackage{ragged2e}
\usepackage{graphicx}
\usepackage{caption}
\usepackage{fancyhdr}
\usepackage{gensymb}

\begin{acronym}
\acro{qcl}[QCL]{Quantum cascade laser}
\acro{icl}[ICL]{Interband cascade laser}
\acro{mir}[MIR]{mid-infrared}
\acro{hQW}[h-QW]{hole-quantum well}
\acro{liv}[LIV]{light-current-voltage}

\end{acronym}

\begin{document}

\newcommand{\D}{\mathrm{d}}
\newcommand{\kp}{k$\cdot$p}
\newcommand{\Jth}{J_\mathrm{th}}
\newcommand{\mat}[1]{\mathbf{#1}}
\newcommand{\fk}{{f\mathbf{k}^{\prime}}}
\newcommand{\ik}{{i\mathbf{k}}}
\newcommand{\threesubsection}[1]{
\medskip
\textit{#1}: 
}

\title{Charge Transport in Interband Cascade Lasers: An Ab-Initio Self-Consistent Model}

\author{Andreas Windischhofer}
\email{andreas.windischhofer@tuwien.ac.at}
\author{Nikola Opa$\check{c}$ak}
\author{Benedikt Schwarz}
\email{benedikt.schwarz@tuwien.ac.at}
\affiliation{Institute of Solid State Electronics, TU Wien, Gusshausstrasse 25-25a, 1040 Vienna, Austria}

\date{\today}

\begin{abstract}
\acp{icl} stand out due to their low threshold current and minimal power consumption, rendering them viable sources for compact and mobile devices in the mid-infrared. Since their first demonstration, they experienced major performance improvements. Many of them originate, on one hand, from technological enhancements and, on the other hand, also from restricted numerical analysis. Encouraged by the impact of restricted models, an \ac{icl}-specific simulation tool can lead to performance breakthroughs and a better comprehension of governing mechanisms. Drawing from an evaluation of existing tools designed for quantum cascade structures, we implemented a self-consistent density matrix rate equation model generalized to simulate the transport in both conduction and valence band heterostructures. Albeit the extensive inclusion of the quantum effects, special care was taken to maintain a high numerical efficiency. Our charge transport model additionally considers optical field calculations, allowing for predictive calculations of \ac{liv} curves. We benchmark the model against well-established \ac{icl} designs and demonstrate reliable performance predictability. Additionally, we give detailed insights into device characteristics extracted from our model. This ultimately allows us to deepen our understanding of \acp{icl} and refine existing and generate novel designs. 
\end{abstract}
\maketitle

\justifying
\section{Introduction}
\acresetall
The \ac{mir} spectral region offers a variety of unique applications, such as gas trace sensing, spectroscopy, and telecommunication. Coherent and compact light sources available for \SIrange{3}{10}{\micro\meter} are mostly \acp{qcl} and \acp{icl}. When the demands come to low threshold currents and low power consumption in the \SIrange{3}{5}{\micro\meter} wavelength range then \acp{icl} are the go-to sources \cite{Vurgaftman2013}. Modern applications that prioritize compact and battery-operated devices can largely benefit from the unique characteristics of an \ac{icl}. 

The concept of \acp{icl} was first introduced in the seminal work of Yang~\cite{Yang1995,Yang1996}, around the time of the first \ac{qcl} demonstration~\cite{Faist1994}. Driven by applications, significant work was put into \ac{qcl}-related research over the years, which continuously enabled new performance records. The first demonstration of an \ac{icl} was shown in 1997 at cryogenic temperature~\cite{Kurtz1997}. In the following years, \acp{icl} have garnered increasing attention and undergone significant enhancements, resulting in their current capability to emit several hundred milliwatts of optical power~\cite{Bewley2006,Canedy2005,Kim2008,Vurgaftman2013,Kim2015}.

\acp{icl} incorporate the cascading principle of \acp{qcl}, wherein electrons traverse multiple periodic stages, leading to photon emission via an optical transition within each period. While the optical transition in \acp{qcl} takes place between subbands within the conduction band, the \ac{icl} adopts an interband transition, typically within the so-called W-quantum well (QW)\mbox{\cite{Yang1995,Yang1996}.}

The rapid progress of the \ac{qcl} performance was facilitated by a tight interplay between experimental observations and solid theoretical and numerical models~\cite{Ko2010,Borowik2017,Jirauschek2017,Kolek2022,Jacobs2023}. Over the years, various types of models evolved into regular tools for understanding and optimizing \acp{qcl}. 

On the other side, most of the progress in \acp{icl} emerged from enhanced technology and design adaptations deduced from experimental and phenomenological observations. While some works backed up their observation with theoretical descriptions, they emphasized on secluded parts. Nowadays \acp{icl} profit from important design innovations, such as deploying a W-shaped QW as an active region~\cite{JRMeyer1995} and adapting the doping to rebalance the carriers~\cite{Vurgaftman2011}. A recent fundamental insight in \ac{icl} design demonstrated that intervalence-band absorption represents a major obstacle for \ac{icl} performance~\cite{Knoetig2022}. More importantly, it was shown that the detrimental effects can be mitigated by adjusting the thickness of the GaInSb \ac{hQW} of the W-QW. 

While already restricted models of secluded parts can lead to new designs, a full transport model promises a comprehensive deepening of our understanding of the governing mechanisms in these devices. As a full transport model, we understand a self-consistent approach relying only on known material parameters that is ultimately capable of calculating a characteristic \ac{liv} curve. Such an ab initio model enables predictive calculations. 

In the course of this paper, we initially provide a concise overview of major modeling approaches established in \acp{qcl}, followed by the presentation of the necessary generalization of the model for its application to \acp{icl}. We will discuss encountered challenges and the approaches to tackle them. In order to push the performance of \acp{icl} further, we aimed to develop a model that allows us to gain an understanding of the underlying mechanisms and enable predictive design improvements. The simulation tool is evaluated based on selected experimental results from the literature, where we also gain new insights into key properties of \acp{icl}.

\section{Results}
\subsection{Model description}
\begin{figure*}
	\centering
	\includegraphics[width=\textwidth]{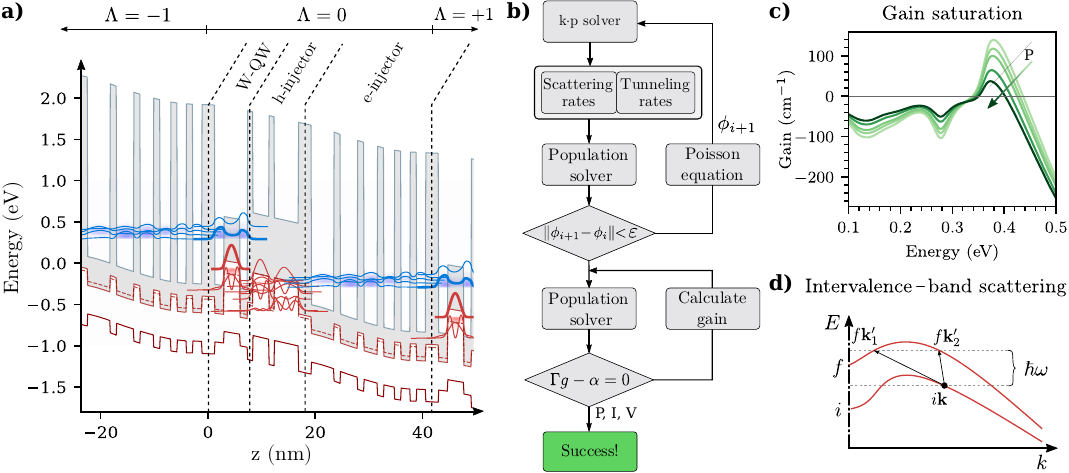}
	\caption{a) Bandstructure of an exemplary \ac{icl}. The displayed heterostructure shows the three sections of an \ac{icl} period, the W-QW, h-injector, and e-injector. Additionally, the first section of the subsequent period $\Lambda=+1$ and the last section of the preceding period $\Lambda=-1$ are shown. The potential landscape includes a linear drop due to an applied bias, and bending from the electrostatic Hartree potential. We display only subbands that are relevant for transport. A red/blue hue represents the relative hole/electron concentration. b) A flowchart of the simulation tool indicating the individual steps and the necessary loops to facilitate self-consistency. c) Exemplary inelastic intervalence-band transition. Due to the non-trivial and possible non-injective nature of the dispersion relation of subbands $E(k)$, rate calculations need to take into account multiple target states $\ik\rightarrow \fk_1, \fk_2$. d) Gain saturation due to the increasing light intensity. The correct output power $P$ is found by increasing the internal photon flux until the gain saturates to zero. }
	\label{fig:intro}
\end{figure*}
In order to choose a suitable model for \acp{icl}, we first evaluate existing models for \acp{qcl}. The development of \acp{qcl} has been significantly pushed since its discovery, owing much to extensive modeling efforts. The approaches to the modeling span from simple and numerically-efficient empirical rate equation models, up to computationally expensive and sophisticated non-equilibrium Green's functions. The twofold intention for our \ac{icl} model is to enable design improvements and to quickly test new innovative ideas. 
This clearly underscores the need for reasonable computation times, while also emphasizing the necessity for a detailed description to accurately predict all relevant mechanisms.

A good compromise between computational effort and accuracy is a semi-classical rate equation approach. In this case, a \ac{qcl} is described as a set of quantized eigenstates, also called subbands, which result from the confining heterostructure potential. The carrier transport is described by rate equations with explicitly calculated transition rates $W_{if}$ from a subband $i$ to $f$ - often also expressed as inverse lifetime $1/\tau_{if}$. In order to find the population of each subband, the summation of all rates - possibly originating from various scattering mechanisms - is fed into a rate equation that collects the currents going out and in of a subband. The rate equations can be written by explicit sums~\cite{QWHarrison2016}
\begin{equation}
	\frac{\D n_i}{\D t} = \sum_{f=0\, , \,f\neq i}^{f=N_\mathrm{sb}-1} \frac{n_f}{\tau_{fi}}-\frac{n_i}{\tau_{if}}
	\label{eqn:rateEquLifetimes}
\end{equation}
or also denoted in a linear matrix representation
\begin{equation}
	\frac{\D}{\D t}  \mathbf n = \mat A \mathbf{n} \mathrm{,}
	\label{eqn:rateEquLifetimesMatrix}
\end{equation}
where $n_i$ is the population of the $i$-th subband and $\mathbf{n}$ a vector of the length $N_\mathrm{sb}$ that collects the populations of all subbands, and $\mat A$ being a matrix whose elements are the inverse lifetimes. In such a semiclassical rate equation model, all rates originate from incoherent scattering mechanisms described by Fermi's golden rule. The Reference~\cite{Jirauschek2014} gives an excellent review of various scattering mechanisms. Extending this approach leads to a microscopic three-dimensional description that includes an explicit $k$-space-resolved dispersion $E_i(k)$~\cite{Iotti2001}. 

An extension of the semi-classical model by means of a density-matrix formalism can also consider quantum mechanical processes such as coherent resonant tunneling~\cite{Iotti2001,Iotti2005}. Compared to the semiclassical rate equation model, the numerical cost of a density-matrix model scales up rapidly, especially when employing $k$-space-resolved subbands. In order to alleviate the numerical cost of the model, a single period of the \ac{qcl} structure can be split into sub-period sections as indicated in Figure \ref{fig:intro}a on the basis of an \ac{icl}~\cite{Terazzi2010}. Within such sections, the charge transport can be solved using self-consistent incoherent rate equations. The transport between neighboring sections is modeled by sequential resonant tunneling~\cite{Terazzi2010,Terazzi2008}. Due to the strict separation between tunneling and incoherent scattering, there are never both mechanisms acting on one transition between two subbands. This allows a reduction of the density matrix problem to a linear rate equation in the form of Equations \ref{eqn:rateEquLifetimes} and \ref{eqn:rateEquLifetimesMatrix}, with a resonant tunneling contribution \cite{Terazzi2010}. By this strategy, the complexity of a full density-matrix formalism can be reduced, while still including selected quantum mechanical effects. We chose this method for its balanced degree of complexity versus efficiency. The main advantage for \acp{icl} is that it allows modeling the semimetallic interface, which constitutes from a special band alignment with the valence band maximum above the conduction band minimum, without the need for mixed valence-conduction band states.~\cite{footnoteSMIF} 
In fact, this chosen approach is actually computationally more efficient than a pure incoherent model, as splitting one cascade into multiple smaller sections leads to a speed-up due to a smaller total number of subband pairs, which outweighs the added costs to calculate tunneling. In the following paragraphs, we will carry the reader briefly through the steps of our model, compare to Figure \ref{fig:intro}b.

In order to calculate a \ac{liv} curve, a term that accounts for the stimulated emission is included in the rate equation $W_{if}^\mathrm{opt} \propto \left| O_{if} \right|^2 P  L(\omega) $. The optical transition rate $W_{if}^\mathrm{opt}$ depends on the intracavity intensity $P$ and the optical matrix element $O_{if}$. For computational efficiency $O_{if}$ can be precalculated after the \kp-solver. For a slightly detuned transition, we consider a lineshape function $L(\omega)$, typically a Lorentzian function. An increasing $P$ leads to gain saturation due to a reduced population inversion. Within a root finder, the population solver runs with adapting $P$ until the condition $\Gamma g-\alpha = 0$ is satisfied, with $\alpha$ being the waveguide losses and $\Gamma$ the confinement factor to the active region. From the spectral picture of the gain saturation, see Figure \ref{fig:intro}c, it shows that the position of the peak gain shifts with increasing $P$, which is further considered in the root finder. 

For \acp{qcl}, a complete transport model can already lead to an extensive numerical cost. Many works simplify certain parts of their code to speed it up, such as neglecting non-parabolicity, Pauli-blocking due to a finite final state population, or a self-consistent Schrödinger-Poisson solution. In the following, we will discuss the consequences of extending typical \ac{qcl} models to \acp{icl}.

\threesubsection{Electronic structure}
The basis of all the following calculations is the electronic structure. The multiband \kp-method combined with the envelope function approximation is a widely used method for modeling semiconductor heterostructures, such as quantum wells, superlattices, and semiconductor lasers~\cite{bastard1988wave}. The number of involved bulk bands used in the applied \kp-method needs to be carefully chosen. To model \acp{qcl}, it is sufficient to use a two-band \kp-model. For large band gap materials, considering conduction and valence bands separately can be a good approximation. On the other side, for narrow band gap materials, such as InAs and InSb, it is advised to consider multiple bands and their coupling~\cite{Kolokolov2003,Jiang2014}. An eight-band \kp-model includes the coupling between the lowest conduction band, the three highest valence bands (heavy holes HH, light holes LH, and split-off SO), and the spin-orbit interaction~\cite{Enders1996}. Within this formalism, strain is considered, so an arbitrary strain configuration still leads to the expected electronic properties.  

The coupling of the bulk bands affects the energy at the $\Gamma$-point as well as the in-plane dispersion relation $E_\parallel(k)$. This also affects the highly relevant type-II W-QW structures, as used as active regions in \acp{icl}, leading to a dispersion relation that is hardly approximated by a constant effective mass. The subbands in the valence band particularly exhibit a $k$-dependent effective mass with values smaller around the $\Gamma$-point than for higher momentum~\cite{Yu2011}. In general, we expect arbitrary forms of dispersion relations.

For the results presented here, we used an exact block-diagonalized four-band \kp-model \cite{Enders1996} with spin-degenerated subbands in order to speed up the calculations. In order to reconstruct the dispersion, we run the \kp-solver for multiple $k$ ($N=\num{16}$) in the range from \SIrange{0}{1.6e9}{\per\meter} and subsequently do a cubic-spline fit for each subband. Instead of an eight-band model~\cite{Foreman1997,VandeWalle1989,Vurgaftman2020}, we opted for the four-band model as this reduces the computational costs. It works in symmetry directions exactly. The approximation is that we use only one spin direction. We samplewise verify that the solution does not differ significantly from the eight-band model. We employ the material parameters based on the valuable Reference~\cite{Vurgaftman2020}. 

In \acp{icl}, the typical populations are significantly larger than in \acp{qcl}, and thus, the impact of the charge distribution onto the electronic potential via the so-called Hartree potential $\phi_\mathrm{Hartree}$ must be considered. It can be obtained from the Poisson equation and is a result of the electrostatic potential due to charge accumulation. Since the charge distribution is not known a priori, a self-consistent Schrödinger-Poisson loop is required, as illustrated in Figure \ref{fig:intro}b. Figure \ref{fig:intro}a indicates the subbands of an \ac{icl} including the Hartree potential that is obtained from consequently solving the transport. 

\threesubsection{Scattering rates}
From Fermi's golden rule we obtain the transition rate $W_{\ik,\fk}$ in a $k$-space resolved form
\begin{equation}
	W_{\ik,\fk} = \frac{2\pi}{\hbar}|\langle \fk | V | \ik \rangle|^2 \delta \left( E_\fk-E_\ik  \pm \hbar\omega \right)
 \label{eqn:FermiGoldenRule}
\end{equation}
where $V$ is the interaction potential and $\ik$ and $\fk$ are the initial and final scattering states. The Dirac function $\delta(E)$ enforces the energy conservation and consequently determines the $\mathbf{k}'$ matching to an initial state $\mathbf{k}$ for a given interaction energy $\hbar\omega$. For parabolic subbands, $\mathbf{k}'$ can be calculated analytically using effective masses~\cite{Jirauschek2014,QWHarrison2016}. In the case of non-parabolic subbands, $\mathbf{k}'$ needs to be numerically computed. We consider LO-phonon and interface roughness scattering within this formalism. For elastic processes, no energy transfer is involved, and therefore $\omega = 0$. For clarity, we omit $\hbar\omega$ in the following discussion of the numerical implementation. 

Our implementation of the \kp-model gives a set of subbands whose dispersion relation is saved as a cubic spline fit. Via an efficient root solver for the cubic splines, we can find the final state $\fk$ given the initial states $\ik$ and the scattering transition energy $\hbar \omega$. Due to the non-injective nature of valence subbands' dispersion $E(k)$, multiple final states can occur. They are all treated as viable individual transitions from $\ik$ to $\fk_n$ with $n$ denoting the $n$-th root of the spline, as illustrated in Figure \ref{fig:intro}d.

In order to decouple the computation of the scattering rates from solving the rate equations, we seek a population-independent form of the scattering rates. For subbands with low carrier density, one can assume Boltzmann statistics within the subband. When Boltzmann statistics are used, an averaged rate ${W_{if}}$ between two subbands can be given without prior knowledge of the population and, therefore, independent from the solution of the rate equation~\cite{Jirauschek2014}. In \acp{qcl}, this assumption is typically a valid approximation and simplifies the numerical implementation. However, for \acp{icl}, we need to use Fermi-Dirac statistics, since we expect significantly higher carrier densities. In order to keep a constant scattering matrix for this case as well, we save the non-averaged, $k$-resolved transition rate. This is required, as the calculation of the transition rates is among the most computationally demanding parts of the model. Any recalculation of rates should be avoided if possible.

For our implementation, we decided to define an equidistant grid in the in-plane wavevector $k$ instead of the energy. An immediate advantage is that all subbands can be discretized using the same grid, independent of their energy. Furthermore, this approach offers inherently higher energy resolution around the $\Gamma$-point, where we expect the majority of the relevant scattering process to happen. Note that, we nevertheless perform the numerical integration of the rates on the resulting non-equidistant energy grid, derived from the $k$ grid, in order to ensure energy conservation. All transition rates are assigned and saved to their $k$ grid point. The following calculations give more details about these essential transformations. 

The current from an initial subband $i$ to a final subband $f$ can be described with averaged rates as $J_{if}=-e(\overline{W_{if}}-\overline{W_{fi}})$ whereas $\overline{W_{if}}$ is obtained as averaged rate from $W_{\ik,\fk}$
\begin{equation}
\overline{W_{if}} = \sum_\mathbf{k} \sum_\mathbf{k'} W_{\ik,\fk} f_i(E_\ik)\left(1-f_f\left(E_\fk\right)\right)
\label{eqn:rateAverage}
\end{equation}
where $f_i$ is the Fermi-Dirac distribution for the subband $i$ with a quasi Fermi energy $E^\mathrm{F}_{i}$ which is omitted in the notation. The term $(1-f_f(E_\fk))$ accounts for Pauli's exclusion principle. 
\begin{figure*}
	\centering
	\includegraphics[width=0.5\textwidth]{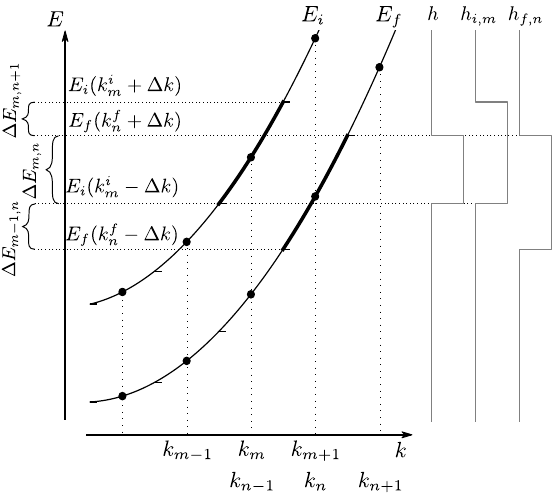}
	\caption{To obtain a scattering rate independent from the population, we calculate and save the scattering rates $k$-space resolved. The discretization in $k$ leads to a grid denoted with an index $m$ or $n$ and a non-equidistant grid in energy. The borders of the integral are limited by the overlap of both window functions $h_{i,m}$ and $h_{f,n}$. For a discretization of the integral for one scattering transition from $m$ to $n$, we take the width $\Delta E_{m,n}$ and evaluate the matrix element at corresponding $k$ values. }
	\label{fig:rateCalc}
\end{figure*}
The sums can be converted to integrals using the density of states \cite{QWHarrison2016,Jirauschek2014}. To this point, we are handling two-dimensional in-plane wave vectors that can be further expressed in polar coordinates, which is especially convenient for symmetric subbands. For a single wave vector, this conversion is written as 
\begin{equation}
    \sum_\mathbf{k} \ldots \rightarrow \frac{2}{(2 \pi)^2} \iint_{-\infty}^{\infty} \ldots \, \D^2 k =  \frac{2}{(2 \pi)^2} \int_0^\infty \int_0^{2\pi} \ldots\, k\D k \D \phi \approx \frac{1}{\pi} \int_0^\infty \ldots\, k\D k =  \frac{1}{\pi} \int_{E(0)}^\infty \ldots\, k \left(\frac{\D E}{\D k}\right)^{-1} \D E \,,
\end{equation}
with a factor \num{2} for the spin degeneracy. The integral $\D \phi$ is resolved by assuming an axial-symmetric subband -- and therefore is constant in $\phi$. This conversion is now applied to Equation \ref{eqn:rateAverage} for final and initial wavevectors, and together with Equation \ref{eqn:FermiGoldenRule} gives 
\begin{equation}
    \overline{W_{if}} = \frac{2}{\pi\hbar} \int_{E_i(0)}^\infty \int_{E_f(0)}^\infty   \underbrace{|\langle \fk| V | \ik \rangle|^2 \, \delta \left( E_\fk-E_\ik \right)f_\ik (1-f_\fk)}_{g(\ik,\fk)} k_i k_f \left(\frac{\D E_i}{\D k_i}\right)^{-1} \left(\frac{\D E_f}{\D k_f}\right)^{-1}  \D E_i \D E_f\,,
\end{equation}
with $k_i$ and $k_f$ describing the magnitude of $\mathbf{k}$ and $\mathbf{k}'$. In the next step, we discretize it on the $k$-grid notated with $k^i_{m}$ ($k^f_{n}$) where $i$ ($f$) stands for the initial (final) subband and $m$ ($n$) are the element indices, 
\begin{equation}
    \overline{W_{if}} = \frac{2}{\pi\hbar} \sum_{m} \int_{E_i(k^i_{m}-\Delta k)}^{E_i(k^i_{m}+\Delta k)} \sum_{n}\int_{E_f(k^f_{n}-\Delta k)}^{E_f(k^f_{n}+\Delta k)}   g(\ik,\fk)\, k_i k_f \left(\frac{\D E_i}{\D k_i}\right)^{-1} \left(\frac{\D E_f}{\D k_f}\right)^{-1} \D E_i \D E_f\,.
\end{equation}
The non-injective nature of the subbands needs to be considered when we evaluate the Dirac function. The summation of multiple final states follows the fundamental principle of the Dirac delta function $\delta(g(x))=\sum_n \frac{1}{|g'(x_n)|}\delta(x-x_n)$. We omit this sum in the calculations for readability; it simply allows us to add the rates for each solution of $\fk_n$. \\
Both integrals with their finite intervals can be rewritten for infinite intervals using a rectangular window function defined as 
\begin{equation}
    h_{i,m}(E_i) = \begin{cases}
        1 &   {E_i(k^i_{m}-\Delta k)} \leq E_i \leq {E_i(k^i_{m}+\Delta k)} \\
        0 & \, \text{else,}
        \end{cases}
\end{equation}
analogously for $f$, as sketched in Figure \ref{fig:rateCalc}. This transformation gives
\begin{equation}
    \overline{W_{if}} = \frac{2}{\pi\hbar} \sum_{m} \int_{-\infty}^{\infty} \sum_{m}\int_{-\infty}^{\infty} h_{i,m}(E_i)h_{f,n}(E_f)  g(\ik,\fk)\, k_i k_f \left(\frac{\D E_i}{\D k_i}\right)^{-1} \left(\frac{\D E_f}{\D k_f}\right)^{-1} \D E_i \D E_f\,.
\end{equation}
In this form, we can easily evaluate the integral $\D E_f$ and the Dirac function, so we get 
\begin{equation}
    \overline{W_{if}} = \frac{2}{\pi\hbar} \sum_{m} \sum_{n} \int_{-\infty}^{\infty} h_{i,m}(E_i)h_{f,n}(E_i)  \tilde{g}(\ik,\fk)\, k_i k_f \left(\frac{\D E_i}{\D k_i}\right)^{-1} \left(\frac{\D E_f}{\D k_f}\right)^{-1} \D E_i \,.
\end{equation}
We can unite the two window functions to a new one $h(E_i) = h_{i,m}(E_i)h_{f,n}(E_i) $ with the width $\Delta E_{m,n}$. For the numerical implementation, we approximate the remaining integral
\begin{equation}
    \overline{W_{if}} \approx \sum_{m} \sum_{n} \underbrace{\frac{2}{\pi\hbar}\Delta E_{m,n}  |\langle fk_{n}^f | V | ik^i_{m} \rangle|^2 k^i_{m}k^f_{n} \left(\frac{\D E_i}{\D k_i}\right)^{-1}\Bigg\rvert_{k_i = k^i_{m}} \left(\frac{\D E_f}{\D k_f}\right)^{-1}\Bigg\rvert_{k_f = k^f_{n}} }_{A_{m,n}} f_i(E_i(k^i_{m})) (1-f_f(E_f(k^f_{m})) )\,.
    \label{eqn:discretizedRate}
\end{equation}
This leaves us with the elements for the rate equation matrix $A_{m,n}$ which are independent of the population.

\threesubsection{Nonlinear rate equation}
The population is obtained by setting up the system of rate equations and solving it in a steady state $\D \mathbf n/\D t = 0$. As discussed above, for \acp{icl} we must use the Fermi-Dirac distribution within the subbands, which makes the scattering rates between subbands dependent on the population. However, a system of equations with constant rates is desirable because it can then be solved separately from the rate calculations.
As the $k$-space rates $W_{\ik,\fk}$ do not depend on the population, we write a $k$-space resolved rate equation in matrix form according to Equation \ref{eqn:discretizedRate}, where the populations for subband $i$ at $k$-point $k_m^i$ are represented with a vector $\mathbf{n}$ of length number of subbands $N_\mathrm{sb}$ times number of $k$-grid points $N_k$, 
\begin{equation}
	\frac{\D}{\D t} \mathbf n = \mathrm{diag}(1-\mathbf f ) \mat A \mathbf f
    \label{eqn:rateEqukSpace}
\end{equation}
where $\mathbf f = f(E_i(k_m^i )-E^\mathrm{F}_{ik}) $ is a vector of same dimension as $\mathbf{n}$, $E_{ik}^\mathrm{F}$ is the Fermi-energy of the state $|ik_m^i\rangle$. Note that the elements of $\mat A$ implicitly contain the density of states. In such a system the scattering matrix $\mat A$ is constant for non-radiative and single electron scattering mechanisms. Stimulated emission and electron-electron scattering to describe the Auger recombination are thus treated separately. 

The rate equations, being a set of ordinary differential equations, are robustly solved for the Fermi-levels $E_{ik}^\mathrm{F}$ by numerical integration from an initial value with an implicit method based on backward-differentiation formulas as described in Reference~\cite{BDF}. The implemented direct nonlinear steady-state solver is computationally more efficient but requires a proper initial condition for reliable operation. We assume that intrasubband scattering effects are sufficiently strong in order to thermalize the subbands, which was realized by adding an artificial intrasubband thermalization rate. Thereby, all $E_{ik}^\mathrm{F}$ within one subband $i$ converge to one value, which consequently allows us to use reduced quasi Fermi energies $E_{i}^\mathrm{F}$ for further calculations.

For both numerical solvers, it is absolutely crucial to utilize an analytic Jacobian of Equation \ref{eqn:rateEqukSpace} with the set of input variables $E_{ik}^\mathrm{F}$ to ensure a stable numerical solution. The Jacobian matrix is commonly used in root solvers and describes a first-order derivative. For general vector-functions $\mathbf{g}(x_n)$ with multiple input variables $x_n$, the elements of the Jacobian matrix are given by $\mathbf{J}_{ij} = \partial g_i / \partial x_j$. A numerically approximated Jacobian is challenged by the Pauli-blocking term $1-f$ when $f$ is close, but not equal, to one. 

So far the effort to keep the scattering matrix for the calculation of the populations constant works only for non-radiative one-body scattering effects. However, intersubband Auger-type processes involve two particles, where one gives up its energy to another particle. Adding this to the rate equation results in a non-linear higher-order problem, that we tackled by adding a reduced term to the non-linear rate equation solver, as well as a proper implementation of the corresponding Jacobian.

\threesubsection{Generalized momentum matrix element}
For the computation of the rate due to interaction with photons, we once again need to carefully evaluate the matrix element for each $k$. This needs to be done with careful consideration of the polarization selection rules since optical transitions in the valence band exhibit nontrivial polarization selection rules~\cite{Vurgaftman2020}. To correctly calculate the gain crosssection we implemented a generalized momentum matrix element model~\cite{Szmulowicz1995,Mu2004,Vurgaftman2020}. This includes contributions from HH, LH, and SO bands. From the nonparabolicity, the optical matrix element is expected to strongly depend on $k$ and is therefore calculated in the $k$-space.

\threesubsection{Emperical dephasing model}
We chose semi-empirical models for a few effects to keep the computational effort reasonably low. The required phenomenological parameters were obtained by adapting the affected characteristics to experimental data, as elaborated in Section \ref{sec:benchmark}.

For the dephasing time $T_2$ of the line broadening function 
\begin{equation}
L(\omega) = \frac{1}{\pi}\frac{1/T_2}{(1/T_2)^2 + (\omega-\omega_\mathrm{ph})^2/\hbar^2} \, ,
\end{equation}
we are using empirical values. For the stimulated rate, we use a Lorentzian function that accounts for line broadening, which is used for all the optical transitions $|\ik\rangle$ to $|\fk\rangle$. A calculation of the dephasing times involves the population and cannot be calculated without its knowledge in the case of highly populated subbands. Therefore, we use empirical dephasing times $T_2$ that are constant for all $k$ values. An adequate fit of the model to the experimental data required at least two different dephasing times depending on whether the transition is interband or intersubband. In accordance with the literature, interband transitions (e.g., CB-HH) exhibit a different broadening than the intersubband transitions (e.g., HH-HH) \cite{Rees1995}. We omitted a separate $T_2$ for CB-CB transitions, as they were not relevant in the energy range of interest. Good values for the interband transitions $T_2^\mathrm{Inter}$ were found to be \SI{100}{\femto\second} and for intraband transitions $T_2^\mathrm{Intra} = \SI{70}{\femto\second}$.

\threesubsection{Emperical Auger model}
Auger recombination is recognized as a significant factor in \acp{icl}, operating as a competing interband transition mechanism that obstructs the lasing transition.
Albeit its importance, a first-principle implementation of Auger recombination is complex and extremely computationally expensive, especially when compared to one-body scattering effects such as phonon-induced scattering \cite{QWHarrison2016}. 
In our model, we implemented the Auger process only for the lasing levels, using a phenomenological model that is benchmarked against experimental data. Thus the Auger rate $W^\mathrm{Auger}_{u,l}$ from the upper $u$ to the lower $l$ lasing level is given by
\begin{equation}
	W^\mathrm{Auger}_{ul} = w_\mathrm{nnp}n^2p + w_\mathrm{npp}np^2
\end{equation}
with $n$ ($p$) being the electron (hole) sheet densities in the upper (lower) lasing level, and $w_\mathrm{nnp}$ ($w_\mathrm{npp}$) as Auger coefficients. In agreement to the literature, we keep the coefficients equal~\cite{Vurgaftman2011}. We found \SI{2e-23}{\centi\meter\tothe{4}\per\second} to give a great match with the experimental data.

\subsection{Model benchmark}
\label{sec:benchmark}
On the basis of two major experimental design studies \cite{Vurgaftman2011,Knoetig2022}, we want to give a qualitative assessment of the model's predictions.

\begin{figure}[ht!]
    \centering
	\includegraphics[]{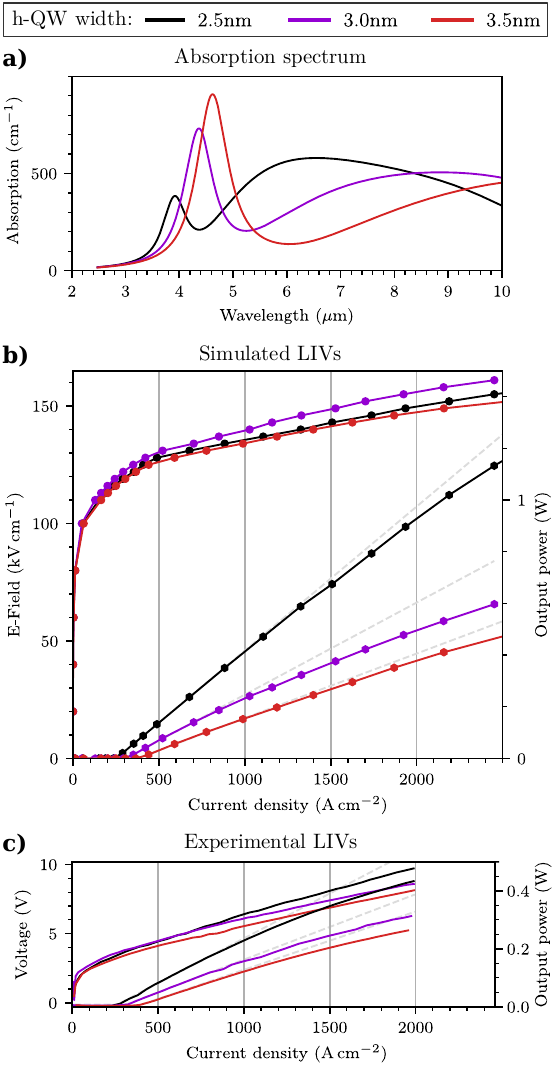}
	\caption{a) Absorption spectra for various \ac{hQW} widths. Changing the \ac{hQW} width in the W-QW causes a change in the absorption spectra. Thereby, the structure can be tuned to have the lowest losses at the target wavelength. At around $\lambda = \SI{4.3}{\micro\meter}$ a device with a \SI{3.5}{\nano\meter} \ac{hQW} has high losses compared to one with \SI{3}{\nano\meter}, while a \SI{2.5}{\nano\meter} \ac{hQW} would have the lowest losses among those designs. b) Simulated \acp{liv} for three \ac{hQW} widths. The LIVs of the designs from \cite{Knoetig2022} were simulated using a density matrix formalism with a self-consistent Schrödinger-Poisson solver. Grey dashed lines are a linear fit close to the threshold to accentuate the efficiency droop. The electric field in \si{\kV\per\centi\meter} can be converted to a voltage \si{\volt} via a conversion factor $\num{0.0175}$ with a limited validity. c) Experimental \acp{liv} for three \ac{hQW} widths. For comparison with the simulation results, we reported experimental data from the supplemental material of Reference~\cite{Knoetig2022} for the \SI{1.8}{\milli\meter} long devices.}
	\label{fig:LIVABC}
\end{figure}
\threesubsection{Intervalence-band absorption}
The authors of Reference~\cite{Knoetig2022} studied the influence of the width of the \ac{hQW} from the active W-QW on typical laser performance parameters, such as threshold current density $\Jth$ and its temperature dependence. By a carefully chosen design to minimize valence intersubband absorption, they enabled a performance boost towards longer wavelengths. This approach was further applied to achieve CW operation of \acp{icl} above \SI{6}{\micro\meter}~\cite{Nauschuetz2023}. 

Our novel simulation tool confirms the essential role of valence intersubband absorption, as seen in Figure~\ref{fig:LIVABC}a. For this calculation, we could access the population given by the solution of the transport model. Note that in Reference~\cite{Knoetig2022}, the authors relied on a best guess for the population of the involved subbands. A reduced internal absorption consequently leads to a reduced $\Jth$. Furthermore, a guideline for designing the ICL active region is provided in terms of choosing the optimal \ac{hQW} width for a desired wavelength, aiming at the lowest losses due to intervalence-band absorption. On the basis of three designs with different thicknesses of the \ac{hQW} -- \SI{2.5}{\nano\meter}, \SI{3}{\nano\meter} and \SI{3.5}{\nano\meter} -- they also showed experimentally the impact of this mitigation strategy.

We were able to simulate the \acp{liv} for all three design variations in Reference~\cite{Knoetig2022}, as seen in Figure~\ref{fig:LIVABC}b. The calculation of a single \ac{liv} with \num{24} discrete bias points, took approximately \SI{25}{\minute} on an Intel\textsuperscript{\textregistered} Core\textsuperscript{TM} i9-10980XE central processing unit. The simulated curves follow the trends of the experimental data, with an excellent match of the threshold current densities. The \SI{2.5}{\nano\meter} thick \ac{hQW} gives $\Jth^\mathrm{Sim}=\SI{251}{\kilo\ampere\per\square\centi\meter}$ from the simulation and $\Jth^\mathrm{Exp} = \SI{247}{\kilo\ampere\per\square\centi\meter}$ from experiment. The \SI{3}{\nano\meter} and \SI{3.5}{\nano\meter} thicknesses compare with $\Jth^\mathrm{Sim} =  \SI{307}{\kilo\ampere\per\square\centi\meter}$ versus $\Jth^\mathrm{Exp} = \SI{310}{\kilo\ampere\per\square\centi\meter}$, and $\Jth^\mathrm{Sim} =  \SI{373}{\kilo\ampere\per\square\centi\meter}$ versus $\Jth^\mathrm{Exp} = \SI{372}{\kilo\ampere\per\square\centi\meter}$, respectively. As it can be seen in Figure \ref{fig:LIVABC}b, the slope efficiency also improves for thinner \ac{hQW}, which is also in accordance with the experimental data, displayed in Figure \ref{fig:LIVABC}c. Our simulation results suggest a higher slope efficiency for all three devices. In reference to the experimental data, the simulation differs by \SI{76}{\percent} for the \SI{2.5}{\nano\meter} device, and \SI{56}{\percent}/\SI{28}{\percent} for the \SI{3.0}{\nano\meter}/\SI{3.5}{\nano\meter} devices, respectively. We see possible reasons for this discrepancy in insecurities of the measurement -- the high divergence of these lasers impedes the collection of the entire emission. 

For the gain saturation, we modeled the Fabry-P\'erot cavity with length \SI{1.8}{\milli\meter}, width \SI{100}{\micro\meter}, given in \cite{Knoetig2022}, a reflectivity of the facets of \SI{32}{\percent} in accordance with \cite{Merritt2015}, modal overlap with the active region of \SI{16}{\percent} \cite{KnoetigDiss2022}, and waveguide losses $\alpha=\SI{270}{\per\centi\meter}$. The estimate for the waveguide losses is based on results from \cite{Knoetig2022}, where the value was extracted by measuring lasers with different cavity lengths. 
We assume the cavity losses to be in the order of the most optimized device, which mitigates intervalence-band absorption.

\begin{figure}[ht!]
	\centering	
	\includegraphics[]{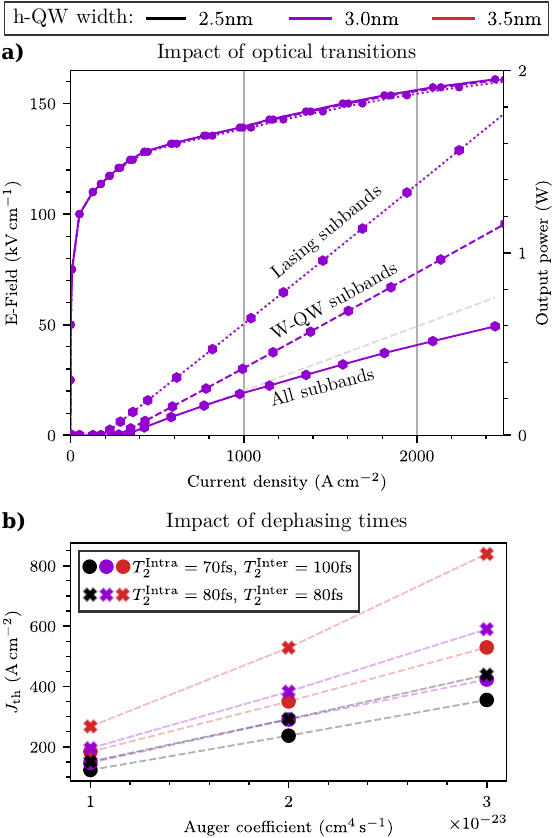}
	\caption{a) Artificially limiting the optical transitions to just the indicated levels suggests their impact on the output power. The inclusion of optical transitions in the h-injector leads to increased losses at higher bias. b) The impact of dephasing times on the current threshold. The sensitivity of the Auger coefficient on the current threshold depends on device design, and also on the dephasing times.}
	\label{fig:ABC_Details}
\end{figure}

In a prior step, we tuned empirical coefficients for a better match. From the Auger coefficient, we expected a linear shift of $\Jth$ with respect to $w_\mathrm{nnp}$. Our simulations reflect this as depicted in Figure \ref{fig:ABC_Details}b. Note that not all devices are equally sensitive to the Auger coefficient. Also, the dephasing times $T_2^\mathrm{Inter} $ for interband transitions, and $T_2^\mathrm{Intra} $ for intraband transitions influence the sensitivity of the threshold current on the Auger coefficient, as seen in Figure~\ref{fig:ABC_Details}b. We carefully tuned the dephasing times to match both the threshold currents and slope efficiency. Searching for a good agreement between the experimental data from \cite{Knoetig2022} and our simulation results, we came to use the following empirical coefficients: $w_\mathrm{nnp}=w_\mathrm{npp}=\SI{2e-23}{\centi\meter\tothe{4}\per\second}$, $T_2^\mathrm{Inter}=\SI{100}{\femto\second}$ and $T_2^\mathrm{Intra}=\SI{70}{\femto\second}$. All data presented in this work uses this set of coefficients and are not modified any further.

Unquestionably, the lasing subbands are by far the most relevant subbands for optical transitions in \acp{icl}. Nevertheless, we show that other optical transitions are of relevance as well. Our simulation tool can give an idea of the impact of various optical transitions. When we only consider the lasing levels for the photon rate, the maximum output is obtained, see Figure~\ref{fig:ABC_Details}a. Further, the light-current relation is strictly linear above the threshold. This does not reflect experimental data, which shows that the efficiency goes down for higher bias currents. Commonly, this effect is known as efficiency droop~\cite{Merritt2015}. Pulsed measurements ensure that this is not a thermal-related effect. By including further transitions within the active W-QW, the output is reduced due to the intervalence-band absorptions in this section. Both the threshold current and slope efficiency deteriorate. Apart from obvious deterioration, we note that the output power increases linearly with current. For a more accurate picture, the intervalence-band transitions within the h-injector also need to be considered. Note that no diagonal optical transitions between the sections (indicated by dashed lines in Figure~\ref{fig:intro}a) are included in this model. The simulation results show an efficiency droop to appear only when intervalence-band absorption within the h-injector is included. When the bias increases, more holes accumulate in the h-injector. Thereby, the absorption losses are increasing, leading to the efficiency droop. 
\begin{figure*}[ht!]
	\centering
	\includegraphics[]{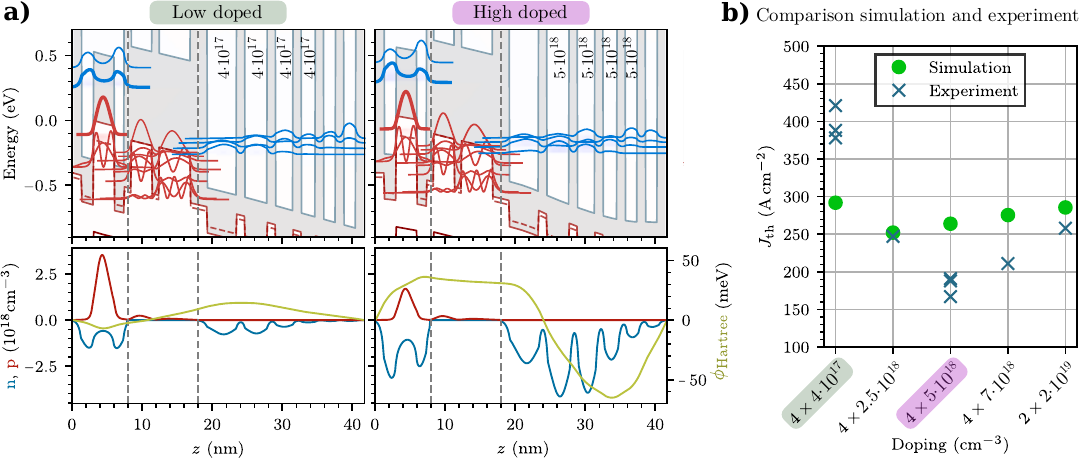}
	\caption{a) Comparison of a low-doped and a high-doped design. A change of the doping concentration from \SI{4e17}{\per\cubic\centi\meter} to  \SI{5e18}{\per\cubic\centi\meter} impacts the ratio of hole to electron concentration in the W-QW. A balanced ratio is expected to lower the threshold currents \cite{Vurgaftman2011}. A higher doping concentration leads to a significant Hartree potential $\phi_\mathrm{Hartree}$ that impacts band alignment. b) Comparison of simulated data and experimental data. The simulation results by our tool of the designs investigated in Reference~\cite{Vurgaftman2011} are compared with the respective experimental data from the same work.}
	\label{fig:Rebalancing}
\end{figure*}

\threesubsection{Carrier rebalancing}
A major breakthrough in \ac{icl} design was made by adapting the doping in the e-injector \cite{Vurgaftman2011}. Based on a quasi-equilibrium distribution calculation, it was shown that earlier devices with low doping have a significantly higher number of holes in the W-QW than electrons. Highlighting the importance of the Auger recombination as a limiting factor, both nnp and npp processes have similar strengths. Therefore, the optimum in terms of $\Jth$ is when both $n$ and $p$ concentrations are approximately balanced. By increasing the doping in the e-injector, a higher electron density is generated in the W-QW. 

Our simulations of the full transport of the structures from \cite{Vurgaftman2011} recreate the experimentally observed trend. In Figure~\ref{fig:Rebalancing}a, we compare two identical designs, only the doping was modified from $4\times\SI{4e17}{\per\cubic\centi\meter}$ to $4\times\SI{5e18}{\per\cubic\centi\meter}$. Both designs are simulated at their respective threshold bias to ensure good comparability. 

Increased doping causes a band bending due to the Hartree electrostatic potential $\phi_\mathrm{Hartree}$, obtained from the Poisson equation. The band structures depict the bending due to the Hartree potential, clearly visible in the e-injector, see Figure~\ref{fig:Rebalancing}a. In the lower subplots of Figure \ref{fig:Rebalancing}a is a direct comparison of $\phi_\mathrm{Hartree}$ is shown beside the carrier densities. In the direct comparison, the shift of the ratio $n$ to $p$ becomes obvious. 

It was experimentally found that a balanced charge carrier density reduces $\Jth$~\cite{Vurgaftman2011}. A comparison of our simulation results with the experimental results from \cite{Vurgaftman2011} shows a similar trend, see Figure~\ref{fig:Rebalancing}b. Although the trend is apparent, the experimental data show a larger impact. We assume, that the empirical Auger model defines the limit here. Instead of the reduced macroscopic model, a detailed microscopic model is expected to lift this limitation~\cite{QWHarrison2016,Gilard1998}.

For all devices of both studies we used the same parameter set. Particularly, we kept the empirical parameters $w_\mathrm{nnp}$, $T_2^\mathrm{Intra}$ and $T_2^\mathrm{Inter}$ constant, despite the fact that the slightly different laser wavelengths of \SI{4.3}{\micro\meter} for Reference~\cite{Knoetig2022} and \SI{3.7}{\micro\meter} in Reference~\cite{Vurgaftman2011} could result in slightly different Auger coefficients, and the epitaxial growth in different reactors could influence the dephasing time. Furthermore, the dephasing times are expected to be affected by the separation of Fermi energies \cite{Rees1995}. 

\section{Conclusion}
In summary, we have pioneered the development of a self-consistent charge transport model for \acp{icl}, employing a first-principle approach. Based on existing models for \acp{qcl}, we followed a density matrix approach with sequential tunneling to model the semimetallic interface. Our endeavor involved careful generalizations to adequately describe transport phenomena within valence subbands. 

For a successful numerical implementation, we made a great effort to keep the scattering matrix in the rate equation independent from populations, which led to a $k$-space resolve rate equation. To facilitate the convergence of the solver, we made use of an implicit solving algorithm for the nonlinear $k$-space resolved rate equations, as well as provided an analytical Jacobian function. For the Auger recombination, we resorted to a macroscopic model with empirical coefficients. The model allows predictive simulations of light-current-voltage characteristics. While this model is capable of predicting trends and describing complex transport phenomena, it is light enough for efficient studying of design improvements and implementation of innovative designs.

In our validation process, we conducted a qualitative comparison of our simulation results with findings from two design studies~\cite{Vurgaftman2011,Knoetig2022}, with good agreement. Despite the few empirical parameters that were needed, we could show that the sensitivity to those parameters is reasonably low. This underscores its efficiency in describing essential transport mechanisms in \acp{icl} and its viability in studying design modifications. Moreover, the evaluation of the model yielded deeper insights into the mechanics of \acp{icl}, such as the importance of intervalence-band absorption in the h-injector and how it induces a significant efficiency droop. This supports that a significant performance boost is possible by reducing internal losses.

Future steps on our road map include replacing the empirical models with exact calculations. We envision that implementing a microscopic Auger model would lead to an advancement in the model's sophistication.

\medskip
\textbf{Acknowledgments} \par 

The authors acknowledge financial support from the European Research Council (ERC) under the European Union’s Horizon 2020 research and innovation programme (Grant agreement No. 853014). This project has been funded with support from the FFG - Austrian Research Promotion Agency and the European Union as part of the Eurostars project “Vaporshine” (No. 904813). Eurostars is part of the European Partnership on Innovative SMEs. The partnership is co-funded by the European Union through Horizon Europe.
We would like to thank Robert Weih and Josephine Nauschütz for their valuable discussions.

\end{document}